\pdfoutput=1

\documentclass[12pt,a4paper]{article}

\usepackage{ifthen} 
\usepackage{subfigure}
\newboolean{pdflatex}
\setboolean{pdflatex}{true} 

\newboolean{articletitles}
\setboolean{articletitles}{true} 

\newboolean{uprightparticles}
\setboolean{uprightparticles}{false} 

\newboolean{inbibliography}
\setboolean{inbibliography}{false} 

\usepackage{microtype}
\usepackage{lineno}  
\usepackage{xspace} 
\usepackage{caption} 

\usepackage{graphicx}  
\usepackage{color}
\usepackage{colortbl}
\graphicspath{{./figs/}} 

\usepackage{amsmath} 
\usepackage{amssymb}
\usepackage{amsfonts}
\usepackage{upgreek} 

\newcommand*\patchAmsMathEnvironmentForLineno[1]{%
\expandafter\let\csname old#1\expandafter\endcsname\csname #1\endcsname
\expandafter\let\csname oldend#1\expandafter\endcsname\csname
end#1\endcsname
 \renewenvironment{#1}%
   {\linenomath\csname old#1\endcsname}%
   {\csname oldend#1\endcsname\endlinenomath}%
}
\newcommand*\patchBothAmsMathEnvironmentsForLineno[1]{%
  \patchAmsMathEnvironmentForLineno{#1}%
  \patchAmsMathEnvironmentForLineno{#1*}%
}
\AtBeginDocument{%
\patchBothAmsMathEnvironmentsForLineno{equation}%
\patchBothAmsMathEnvironmentsForLineno{align}%
\patchBothAmsMathEnvironmentsForLineno{flalign}%
\patchBothAmsMathEnvironmentsForLineno{alignat}%
\patchBothAmsMathEnvironmentsForLineno{gather}%
\patchBothAmsMathEnvironmentsForLineno{multline}%
}

\usepackage{hyperref}    
\usepackage[all]{hypcap} 

\def\lhcb {\mbox{LHCb}\xspace}

\ifthenelse{\boolean{uprightparticles}}%
{

 \def\Ppi         {\ensuremath{\uppi}\xspace}

 \def\PDelta      {\ensuremath{\Delta}\xspace}                 
 \def\PXi      {\ensuremath{\Xi}\xspace}                 
 \def\PLambda      {\ensuremath{\Lambda}\xspace}                 
 \def\PSigma      {\ensuremath{\Sigma}\xspace}                 
 \def\POmega      {\ensuremath{\Omega}\xspace}                 
 \def\PUpsilon      {\ensuremath{\Upsilon}\xspace}                 
 
 \def\PB      {\ensuremath{\mathrm{B}}\xspace}                 
                  
 \def\PD      {\ensuremath{\mathrm{D}}\xspace}

 \def\PK      {\ensuremath{\mathrm{K}}\xspace}

 \def\Pb      {\ensuremath{\mathrm{b}}\xspace}                 
 \def\Pc      {\ensuremath{\mathrm{c}}\xspace}

 \def\Pi      {\ensuremath{\mathrm{i}}\xspace}

 \def\Ps      {\ensuremath{\mathrm{s}}\xspace}

}
{

 \def\Ppi         {\ensuremath{\pi}\xspace}

 \mathchardef\PDelta="7101
 \mathchardef\PXi="7104
 \mathchardef\PLambda="7103
 \mathchardef\PSigma="7106
 \mathchardef\POmega="710A
 \mathchardef\PUpsilon="7107
                  
 \def\PB      {\ensuremath{B}\xspace}                 
                  
 \def\PD      {\ensuremath{D}\xspace}

 \def\PK      {\ensuremath{K}\xspace}

 \def\Pb      {\ensuremath{b}\xspace}                 
 \def\Pc      {\ensuremath{c}\xspace}

 \def\Pi      {\ensuremath{i}\xspace}

 \def\Ps      {\ensuremath{s}\xspace}

}

\def\squark    {\ensuremath{\Ps}\xspace}

\def\cquark    {\ensuremath{\Pc}\xspace}

\def\bquark    {\ensuremath{\Pb}\xspace}

\def\pion  {\ensuremath{\Ppi}\xspace}

\def\pip   {\ensuremath{\pion^+}\xspace}
\def\pim   {\ensuremath{\pion^-}\xspace}
\def\pipm  {\ensuremath{\pion^\pm}\xspace}

\def\kaon  {\ensuremath{\PK}\xspace}
  \def\Kbar  {\kern 0.2em\overline{\kern -0.2em \PK}{}\xspace}

\def\Kz    {\ensuremath{\kaon^0}\xspace}
\def\Kzb   {\ensuremath{\Kbar^0}\xspace}
\def\Kp    {\ensuremath{\kaon^+}\xspace}
\def\Km    {\ensuremath{\kaon^-}\xspace}
\def\Kpm   {\ensuremath{\kaon^\pm}\xspace}
\def\Kmp   {\ensuremath{\kaon^\mp}\xspace}
\def\KS    {\ensuremath{\kaon^0_{\rm\scriptscriptstyle S}}\xspace}

  \def\Dbar    {\kern 0.2em\overline{\kern -0.2em \PD}{}\xspace}
\def\D       {\ensuremath{\PD}\xspace}

\def\Dz      {\ensuremath{\D^0}\xspace}
\def\Dzb     {\ensuremath{\Dbar^0}\xspace}
\def\Dp      {\ensuremath{\D^+}\xspace}
\def\Dm      {\ensuremath{\D^-}\xspace}
\def\Dpm     {\ensuremath{\D^\pm}\xspace}

\def\Dspm    {\ensuremath{\D^{\pm}_\squark}\xspace}

\def\Bbar    {\ensuremath{\kern 0.18em\overline{\kern -0.18em \PB}{}}\xspace}

  \def\Y#1S{\ensuremath{\PUpsilon{(#1S)}}\xspace}

\def\Lz {\ensuremath{\PLambda}\xspace}
\def\Lbar {\ensuremath{\kern 0.1em\overline{\kern -0.1em\PLambda}}\xspace}

\def\BF         {{\ensuremath{\cal B}\xspace}}

\def\BR         {\BF}

\def\to                 {\ensuremath{\rightarrow}\xspace}

\def\CP                {\ensuremath{C\!P}\xspace}

\newcommand{\ACP}{\ensuremath{{\cal A}^{\CP}}\xspace}

\def\AT#1     {\ensuremath{A_{\mathrm{T}}^{#1}}\xspace}           

\def\C#1      {\ensuremath{\mathcal{C}_{#1}}\xspace}                       
\def\Cp#1     {\ensuremath{\mathcal{C}_{#1}^{'}}\xspace}                    
\def\Ceff#1   {\ensuremath{\mathcal{C}_{#1}^{\mathrm{(eff)}}}\xspace}        
\def\Cpeff#1  {\ensuremath{\mathcal{C}_{#1}^{'\mathrm{(eff)}}}\xspace}       
\def\Ope#1    {\ensuremath{\mathcal{O}_{#1}}\xspace}                       
\def\Opep#1   {\ensuremath{\mathcal{O}_{#1}^{'}}\xspace}                    

\newcommand{\tev}{\ifthenelse{\boolean{inbibliography}}{\ensuremath{~T\kern -0.05em eV}\xspace}{\ensuremath{\mathrm{\,Te\kern -0.1em V}}\xspace}}
\newcommand{\gev}{\ensuremath{\mathrm{\,Ge\kern -0.1em V}}\xspace}
\newcommand{\mev}{\ensuremath{\mathrm{\,Me\kern -0.1em V}}\xspace}
\newcommand{\kev}{\ensuremath{\mathrm{\,ke\kern -0.1em V}}\xspace}
\newcommand{\ev}{\ensuremath{\mathrm{\,e\kern -0.1em V}}\xspace}
\newcommand{\gevc}{\ensuremath{{\mathrm{\,Ge\kern -0.1em V\!/}c}}\xspace}
\newcommand{\mevc}{\ensuremath{{\mathrm{\,Me\kern -0.1em V\!/}c}}\xspace}
\newcommand{\gevcc}{\ensuremath{{\mathrm{\,Ge\kern -0.1em V\!/}c^2}}\xspace}
\newcommand{\gevgevcccc}{\ensuremath{{\mathrm{\,Ge\kern -0.1em V^2\!/}c^4}}\xspace}
\newcommand{\mevcc}{\ensuremath{{\mathrm{\,Me\kern -0.1em V\!/}c^2}}\xspace}

\def\mum  {\ensuremath{{\,\upmu\rm m}}\xspace}

\def\invfb   {\ensuremath{\mbox{\,fb}^{-1}}\xspace}

\newcommand{\chisq}{\ensuremath{\chi^2}\xspace}

\newcommand{\chisqip}{\ensuremath{\chi^2_{\rm IP}}\xspace}

\def\gsim{{~\raise.15em\hbox{$>$}\kern-.85em
          \lower.35em\hbox{$\sim$}~}\xspace}
\def\lsim{{~\raise.15em\hbox{$<$}\kern-.85em
          \lower.35em\hbox{$\sim$}~}\xspace}

\def\pt         {\mbox{$p_{\rm T}$}\xspace}

\def\evtgen     {\mbox{\textsc{EvtGen}}\xspace}

\def\geant      {\mbox{\textsc{Geant4}}\xspace}

\def\photos     {\mbox{\textsc{Photos}}\xspace}

\def\pythia     {\mbox{\textsc{Pythia}}\xspace}

\def\tell1  {TELL1\xspace}
\def\ukl1   {UKL1\xspace}

\usepackage{cite} 
\usepackage{mciteplus}

\usepackage{longtable} 

\newcommand{\Reference}{Ref.}
\newcommand{\References}{Refs.}

\newcommand{\Table}{Table}

\newcommand{\Equation}{Eq.}
\newcommand{\Equations}{Eqs.}

\newcommand{\Dcand}{\ensuremath{D^{\pm}_{(s)}}}

\newcommand{\KsToPiPi}{\ensuremath{\KS\to\pip\pim}}

\newcommand{\DToKsh}{\ensuremath{\Dpm_{(s)}\rightarrow \KS h^{\pm}}}
\newcommand{\DToKshprime}{\ensuremath{\Dpm_{(s)}\rightarrow \KS h^{\prime\pm}}}
\newcommand{\DandDsToKsPi}{\ensuremath{\Dpm_{(s)}\rightarrow \KS \pi^{\pm}}}
\newcommand{\DandDsToKsK}{\ensuremath{\Dpm_{(s)}\rightarrow \KS \kaon^{\pm}}}

\newcommand{\DpToKshp}{\ensuremath{\Dp_{(s)}\rightarrow \KS h^+}}
\newcommand{\DmToKshm}{\ensuremath{\Dm_{(s)}\rightarrow \KS h^-}}

\newcommand{\DcandToKsPi}{\ensuremath{\Dcand \rightarrow \KS \pipm}}
\newcommand{\DcandToKsK}{\ensuremath{\Dcand \rightarrow \KS \Kpm}}

\newcommand{\DcandToKsPip}{\ensuremath{\Dp_{(s)} \rightarrow \KS \pip}}
\newcommand{\DcandToKsKp}{\ensuremath{\Dp_{(s)} \rightarrow \KS \Kp}}
\newcommand{\DcandToKsPim}{\ensuremath{\Dm_{(s)} \rightarrow \KS \pim}}
\newcommand{\DcandToKsKm}{\ensuremath{\Dm_{(s)} \rightarrow \KS \Km}}

\newcommand{\DcandToPhiPip}{\ensuremath{\Dp_{(s)} \rightarrow \phi \pip}}
\newcommand{\DcandToPhiPim}{\ensuremath{\Dm_{(s)} \rightarrow \phi \pim}}

\newcommand{\DToKsPi}{\ensuremath{\Dpm \rightarrow \KS \pipm}}

\newcommand{\DToKsK}{\ensuremath{\Dpm \rightarrow \KS \Kpm}}

\newcommand{\DsToKsPi}{\ensuremath{\Dspm \rightarrow \KS \pipm}}

\newcommand{\DsToKsK}{\ensuremath{\Dspm \rightarrow \KS \Kpm}}

\newcommand{\DToKPiPi}{\ensuremath{\Dpm \rightarrow \Kmp \pipm \pipm}}

\newcommand{\DzToKpKm}{\ensuremath{\Dz \rightarrow \Kp \Km}}
\newcommand{\DzToPipPim}{\ensuremath{\Dz \rightarrow \pip \pim}}

\newcommand{\DToPhiPi}{\ensuremath{\Dpm \rightarrow \phi \pipm }}

\newcommand{\DsToPhiPi}{\ensuremath{\Dspm \rightarrow \phi \pipm }}

\newcommand{\DandDsToPhiPi}{\ensuremath{\Dpm_{(s)}  \rightarrow \phi \pipm }}

\renewcommand{\ACP}{\ensuremath{\mathcal{A}_{\CP}}}

\newcommand{\ACPmeas}{\mathcal{A}_{\mathrm{meas}}}
\newcommand{\ACPprod}{\mathcal{A}_{\mathrm{prod}}}
\newcommand{\ACPdet}{\mathcal{A}_{\mathrm{det}}}

\newcommand{\ACPKzKzb}{\mathcal{A}_{\Kz/\Kzb}}
\newcommand{\ACPKz}{\mathcal{A}_{\Kz}}
\newcommand{\ACPKzb}{\mathcal{A}_{\Kzb}}
\newcommand{\ACPDD}{\mathcal{A}_{\CP}^{\mathcal{DD}}}

\usepackage{cite} 
\usepackage{mciteplus}

\begin{document}

\renewcommand{\thefootnote}{\fnsymbol{footnote}}
\setcounter{footnote}{1}


\begin{titlepage}
\pagenumbering{roman}

\vspace*{-1.5cm}
\centerline{\large EUROPEAN ORGANIZATION FOR NUCLEAR RESEARCH (CERN)}
\vspace*{1.5cm}
\hspace*{-0.5cm}
\begin{tabular*}{\linewidth}{lc@{\extracolsep{\fill}}r}
\ifthenelse{\boolean{pdflatex}}
{\vspace*{-2.7cm}\mbox{\!\!\!\includegraphics[width=.14\textwidth]{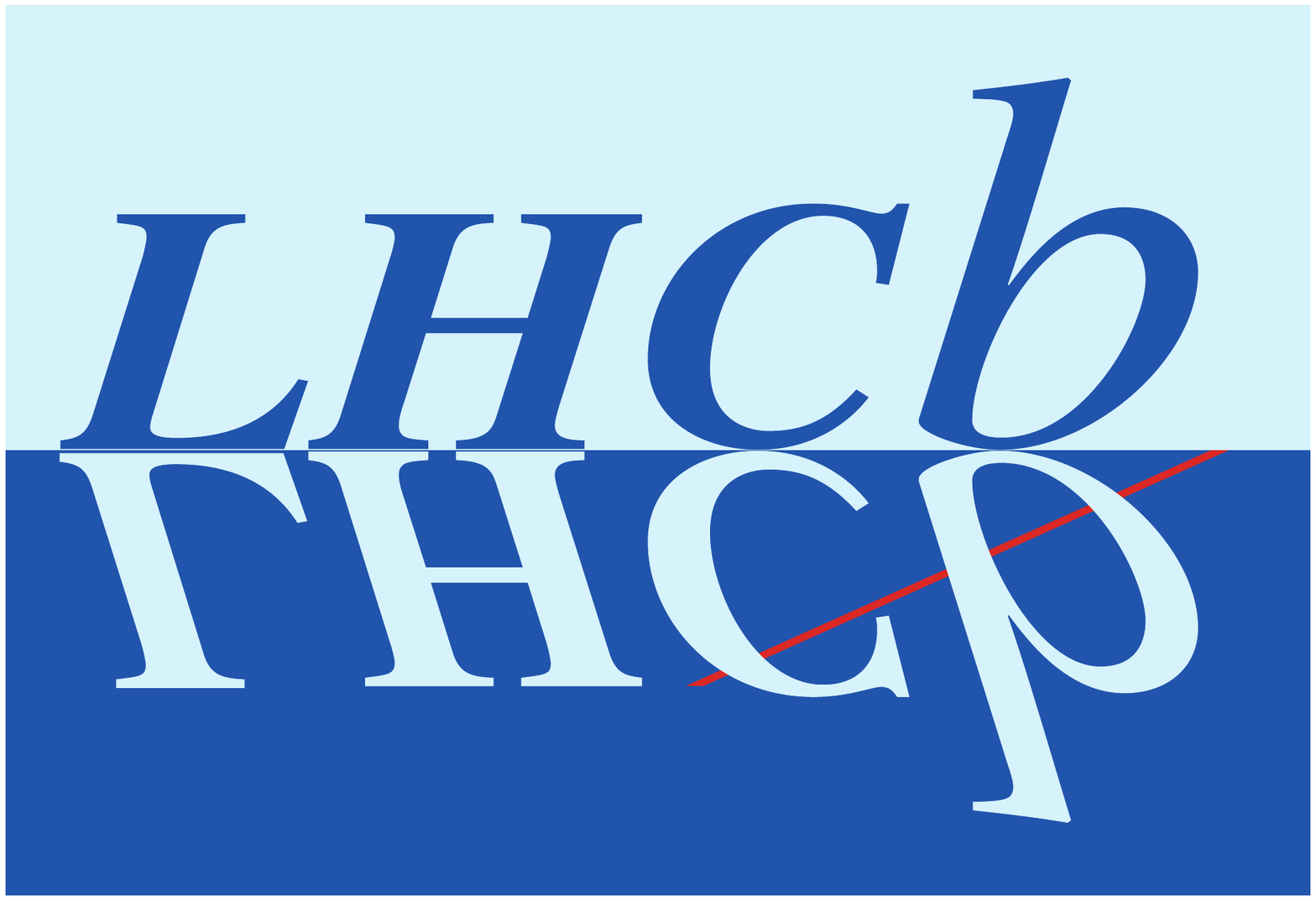}} & &}%
{\vspace*{-1.2cm}\mbox{\!\!\!\includegraphics[width=.12\textwidth]{lhcb-logo.eps}} & &}%
\\
 & & CERN-PH-EP-2014-125 \\  
 & & LHCb-PAPER-2014-018 \\  
 & & 10 June 2014  \\ 
 & & \\
\end{tabular*}

\vspace*{2.0cm}

{\bf\boldmath\huge
\begin{center}
  Search for $\boldmath{\CP}$ violation in 
  $D^{\pm}\rightarrow \KS K^{\pm}$ and
  $D^{\pm}_{s}\rightarrow \KS \pi^{\pm}$ decays
\end{center}
}

\vspace*{1.0cm}

\begin{center}
The LHCb collaboration\footnote{Authors are listed on the following pages.}
\end{center}

\vspace{\fill}

\begin{abstract}  
  \noindent
  A search for \CP violation in Cabibbo-suppressed
  $D^{\pm}\rightarrow \KS K^{\pm}$ and 
  $D^{\pm}_{s}\rightarrow \KS \pi^{\pm}$ decays 
  is performed using $pp$ collision data, 
  corresponding to an integrated luminosity of 3\invfb,
  recorded by the LHCb experiment.
  The individual \CP-violating asymmetries
  are measured to be
\begin{eqnarray*}
  \ACP^{\DToKsK} & = &  (+0.03 \pm 0.17 \pm 0.14) \% \\  
  \ACP^{\DsToKsPi} & = & (+0.38 \pm 0.46 \pm 0.17) \%,
\end{eqnarray*}
  assuming that \CP\ violation in the Cabibbo-favoured decays is negligible.
  A combination of the measured asymmetries for the four decay modes 
  $D^{\pm}_{(s)}\rightarrow \KS K^{\pm}$ and 
  $D^{\pm}_{(s)}\rightarrow \KS \pi^{\pm}$ 
  gives the sum
\[
  \ACP^{\DToKsK} + \ACP^{\DsToKsPi} = (+0.41 \pm 0.49 \pm 0.26) \%.
\]
  In all cases, 
  the first uncertainties are statistical and the second systematic. 
  The results represent the most precise measurements of these 
  asymmetries to date and show no evidence for \CP violation.
  
\end{abstract}

\vspace*{0.5cm}

\begin{center}
  Submitted to JHEP
\end{center}

\vspace{\fill}

{\footnotesize 
\centerline{\copyright~CERN on behalf of the \lhcb collaboration, license \href{http://creativecommons.org/licenses/by/3.0/}{CC-BY-3.0}.}}
\vspace*{2mm}

\end{titlepage}


\newpage
\setcounter{page}{2}
\mbox{~}
\newpage

\centerline{\large\bf LHCb collaboration}
\begin{flushleft}
\small
R.~Aaij$^{41}$, 
B.~Adeva$^{37}$, 
M.~Adinolfi$^{46}$, 
A.~Affolder$^{52}$, 
Z.~Ajaltouni$^{5}$, 
S.~Akar$^{6}$, 
J.~Albrecht$^{9}$, 
F.~Alessio$^{38}$, 
M.~Alexander$^{51}$, 
S.~Ali$^{41}$
% ...

{\footnotesize \it
% ...
$ ^{5}$Clermont Universit\'{e}, Universit\'{e} Blaise Pascal, CNRS/IN2P3, LPC, Clermont-Ferrand, France\\
$ ^{6}$CPPM, Aix-Marseille Universit\'{e}, CNRS/IN2P3, Marseille, France\\
% ...
$ ^{9}$Fakult\"{a}t Physik, Technische Universit\"{a}t Dortmund, Dortmund, Germany\\
% ...
$ ^{37}$Universidad de Santiago de Compostela, Santiago de Compostela, Spain\\
$ ^{38}$European Organization for Nuclear Research (CERN), Geneva, Switzerland\\
% ...
$ ^{41}$Nikhef National Institute for Subatomic Physics, Amsterdam, The Netherlands\\
% ...
$ ^{46}$H.H. Wills Physics Laboratory, University of Bristol, Bristol, United Kingdom\\
% ...
$ ^{51}$School of Physics and Astronomy, University of Glasgow, Glasgow, United Kingdom\\
$ ^{52}$Oliver Lodge Laboratory, University of Liverpool, Liverpool, United Kingdom\\
% ...
}
\end{flushleft}

\cleardoublepage


\renewcommand{\thefootnote}{\arabic{footnote}}
\setcounter{footnote}{0}



\pagestyle{plain} 
\setcounter{page}{1}
\pagenumbering{arabic}


%


\section{Introduction}
\label{sec:Introduction}

Measurements of \CP violation in charm meson 
decays offer a unique opportunity
to search for physics beyond the Standard Model (SM).
In the SM, \CP violation in the charm sector
is expected to be \ensuremath{\mathcal{O}\left( 0.1\% \right) } 
or below~\cite{Bianco:2003vb}.
Any enhancement would be an indication of physics beyond the SM.
Recent measurements of the difference in \CP asymmetries between
\DzToKpKm\ and \DzToPipPim\ decays by the
LHCb~\cite{LHCb-CONF-2013-003,LHCb-PAPER-2013-003,LHCb-PAPER-2014-013},
CDF~\cite{Collaboration:2012qw}, 
Belle~\cite{Ko:2012px} and BaBar~\cite{Aubert:2007if} collaborations 
are consistent with SM expectations, 
although initial results from 
LHCb indicated otherwise~\cite{LHCb-PAPER-2011-023}.
Further investigations in other charm decay modes
are therefore important to provide a more complete 
picture of \CP violation in the charm sector.

In this paper, \CP violation in singly Cabibbo-suppressed 
\DToKsK\ and \DsToKsPi\ decays is investigated.
In the SM, the magnitude of \CP violation in these decays 
is expected to be small, \ensuremath{\mathcal{O} \left( 10^{-4} \right)}, 
excluding the known contribution from \Kz\ mixing~\cite{Lipkin:1999qz}. 
If processes beyond the SM contain additional weak phases,
other than those contained in the Cabibbo-Kobayashi-Maskawa formalism,
additional \CP-violating effects 
could arise~\cite{Lipkin:1999qz, Bhattacharya:2012ah}.

Several searches for \CP violation in \DToKsK\ and \DsToKsPi\ 
decays have been performed 
previously~\cite{Link:2001zj,Mendez:2009aa,Ko:2010ng,Lees:2012jv,Ko:2012uh,LHCb-PAPER-2012-052}.
The \CP asymmetry for \DToKsh\ decays is defined as
\begin{align}
\ACP^{\DToKsh} \equiv \dfrac
{\Gamma(\DpToKshp)- \Gamma(\DmToKshm)}
{\Gamma(\DpToKshp)+ \Gamma(\DmToKshm)},
\end{align} 
where $h$ is a pion or kaon and 
\ensuremath{\Gamma} is the partial decay width. 
The most precise measurements of the \CP asymmetries 
in the decay modes
\DToKsK\ and \DsToKsPi\
are $\ACP^{\DToKsK} = (-0.25 \pm 0.31) \%$ 
from the Belle collaboration~\cite{Ko:2012uh}
and $\ACP^{\DsToKsPi} = (+0.61 \pm 0.84) \%$ 
from the LHCb collaboration~\cite{LHCb-PAPER-2012-052},
respectively.
Both measurements are consistent with \CP symmetry.  
The measurement of $\ACP^{\DsToKsPi}$ by LHCb~\cite{LHCb-PAPER-2012-052} 
was performed using data 
corresponding to an integrated luminosity of 1\invfb, 
and is superseded by the result presented here.
    
In this paper, the \CP asymmetries are determined 
from the measured asymmetries,
\begin{align}
\label{e:ACPsignal}
\ACPmeas^{\DToKsh} = \dfrac
{N_{\mathrm{sig}}^{\DpToKshp}-N_{\mathrm{sig}}^{\DmToKshm}}
{N_{\mathrm{sig}}^{\DpToKshp}+N_{\mathrm{sig}}^{\DmToKshm}},
\end{align}
where $N_{\mathrm{sig}}^{\DToKsh}$ is the signal yield in the decay
mode \DToKsh.
The measured asymmetries include 
additional contributions other than $\ACP^{\DToKsh}$,
such that, 
when the considered asymmetries are small,
it is possible to approximate
\begin{align}
\label{e:ACPDToKsh}
\ACPmeas^{\DToKsh}
\approx \ACP^{\DToKsh} + \ACPprod^{\Dpm_{(s)}}+ \ACPdet^{h^{\pm}} +\ACPKzKzb,
\end{align}
where
\ensuremath{\ACPprod^{\Dp_{(s)}}} is 
the asymmetry in the production of 
\ensuremath{\D^{\pm}_{(s)}} mesons in high-energy $pp$ collisions
in the forward region,
and
\ensuremath{\ACPdet^{h^+}} arises from
the difference in detection efficiencies
between positively and negatively charged hadrons.
The asymmetry  
\ensuremath{\ACPKz \equiv (N_{\Kz}-N_{\Kzb})/(N_{\Kz}+N_{\Kzb})= - \ACPKzb},
where $N_{\Kz/\Kzb}$ is the number of $\Kz/\Kzb$ mesons produced, 
takes into account the detection asymmetry 
between a $\Kz$ and a $\Kzb$ meson due to regeneration and 
the presence of mixing and \CP violation in the $\Kz$-$\Kzb$ system. 
The contribution from the neutral kaon asymmetries 
is estimated using the method described 
in \Reference~\cite{LHCb-PAPER-2014-013}
and the reconstructed \DToKsh\ candidates selected in this analysis.
The result 
\ensuremath{\ACPKz = (+0.07\pm 0.02)\%}
is included as a correction to the measured asymmetries as shown below.  

The \ensuremath{\D^{\pm}_{(s)}} production and hadron detection asymmetries 
approximately cancel by constructing a
{\it double difference} (\ensuremath{\mathcal{DD}})
between the four measured asymmetries, 
\begin{align}
\label{e:ACPDD}
\ACPDD 
& = \left[\ACPmeas^{\DsToKsPi}-\ACPmeas^{\DsToKsK}\right]-
\left[\ACPmeas^{\DToKsPi}-\ACPmeas^{\DToKsK}\right]- 2\ACPKz.
\end{align}
Assuming that \CP violation in the 
Cabibbo-favoured decays is negligible,
\ensuremath{\ACPDD} is a measurement of the sum of the \CP-violating
asymmetries in \DToKsK\ and \DsToKsPi\ decays,
\begin{equation}
\label{e:ACPDDest}
\ACP^{\DToKsK} + \ACP^{\DsToKsPi} = \ACPDD.
\end{equation}
The quantity \ensuremath{\ACPDD} provides a measurement that is largely
insensitive to production and instrumental asymmetries,
even though the \CP asymmetries in \DToKsK\ and \DsToKsPi\ decays are expected
to have the opposite sign. 
 
The individual \CP asymmetries for \DToKsK\ and \DsToKsPi\ decays 
are also determined using the asymmetry measured in the Cabibbo-favoured
decay \DsToPhiPi,
\begin{equation}
\label{e:ACPDToKsK}
\ACP^{\DToKsK} 
= 
\left[\ACPmeas^{\DToKsK}-\ACPmeas^{\DsToKsK}\right]-
\left[\ACPmeas^{\DToKsPi}-\ACPmeas^{\DsToPhiPi}\right]-\ACPKz
\end{equation}
and
\begin{equation}
\label{e:ACPDsToKsPi}
\ACP^{\DsToKsPi} 
= 
\ACPmeas^{\DsToKsPi}-\ACPmeas^{\DsToPhiPi}-\ACPKz.
\end{equation}
Measurements of the sum
\ensuremath{\ACP^{\DToKsK}+\ACP^{\DsToKsPi}},
and the individual \CP asymmetries,
\ensuremath{\ACP^{\DToKsK}} and \ensuremath{\ACP^{\DsToKsPi}}, 
are presented in this paper.

\section{Detector and software}
\label{sec:Detector}

The \lhcb detector~\cite{Alves:2008zz} is a single-arm forward
spectrometer covering the \mbox{pseudorapidity} range $2<\eta <5$,
designed for the study of particles containing \bquark or \cquark
quarks. The detector includes a high-precision tracking system
consisting of a silicon-strip vertex detector surrounding the $pp$
interaction region, a large-area silicon-strip detector located
upstream of a dipole magnet with a bending power of about
$4{\rm\,Tm}$, and three stations of silicon-strip detectors and straw
drift tubes placed downstream.
The polarity of the dipole magnet is reversed 
periodically throughout data-taking.
The combined tracking system provides a momentum measurement with
relative uncertainty that varies from 0.4\% at 5\gevc to 0.6\% at 100\gevc,
and impact parameter resolution of 20\mum for
tracks with large transverse momentum, \pt. 
Different types of charged hadrons are distinguished by information
from two ring-imaging Cherenkov (RICH) detectors~\cite{LHCb-DP-2012-003}. 
Photon, electron and
hadron candidates are identified by a calorimeter system consisting of
scintillating-pad and preshower detectors, an electromagnetic
calorimeter and a hadronic calorimeter. Muons are identified by a
system composed of alternating layers of iron and multiwire
proportional chambers~\cite{LHCb-DP-2012-002}.
The trigger~\cite{LHCb-DP-2012-004} consists of a
hardware stage, based on information from the calorimeter and muon
systems, an inclusive software stage, which uses the tracking system,
and a second software stage that exploits the full event
reconstruction.

The data used in this analysis corresponds to an integrated
luminosity of approximately 3\invfb recorded in $pp$ collisions at 
centre-of-mass energies of $\sqrt{s}=7$\tev ($1$\invfb) 
and 8\tev ($2$\invfb).
Approximately 50\% of the data were collected in 
each configuration ({\it Up} and {\it Down}) of the magnet polarity.

In the simulation, $pp$ collisions are generated using
\pythia~6.4~\cite{Sjostrand:2006za} with a specific \lhcb
configuration~\cite{LHCb-PROC-2010-056}.  Decays of hadronic particles
are described by \evtgen~\cite{Lange:2001uf}, in which final state
radiation is generated using \photos~\cite{Golonka:2005pn}. The
interaction of the generated particles with the detector and its
response are implemented using the \geant
toolkit~\cite{Allison:2006ve, *Agostinelli:2002hh} as described in
Ref.~\cite{LHCb-PROC-2011-006}.

\section{Candidate selection}
\label{sec:Selection}

Candidate \DToKsh\ and \DandDsToPhiPi\ decays are reconstructed
from combinations of charged particles that are well-measured,
have information in all tracking detectors
and are identified as either a pion or kaon, but not as an electron or muon.
The primary $pp$ interaction vertex (PV) is chosen to be the one 
yielding the minimum \chisqip of the \Dcand\ meson,
where \chisqip is defined as the
difference in \chisq of a given PV reconstructed with and
without the considered particle.
The \chisqip requirements discussed below are defined with respect to all PVs
in the event. 

Candidate \DToKsh\ decays are reconstructed from a \KsToPiPi\ decay 
candidate combined with a charged (bachelor) hadron.
The bachelor hadron is required to have $p>5$\gevc,
$\pt>0.5$\gevc
and is classified as a pion or kaon according to
the RICH particle identification information.
The \KS candidate is formed from a pair of 
oppositely charged particles, 
which have $p>2$\gevc, $\pt>0.25$\gevc, $\chisqip>40$,
and are identified as pions.
The \KS is also required to have a good quality vertex fit,
$\pt>1$\gevc, $\chisqip>7$, 
a decay vertex separated from the PV 
by a distance greater than $20$~mm, 
as projected on to the beam direction,
and have a flight distance
$\chi^2>300$.
The \KS mass is constrained to its known value~\cite{PDG2012} 
when the decay vertex is formed and the \Dcand\ mass calculated.
The electron and muon particle identification, 
flight distance and impact parameter
requirements on the \KS reduce backgrounds from
semileptonic 
$\Dcand\to\KS\ell^{\pm}\bar{\nu}_{\ell}$ ($\ell= e$ or $\mu$)
and $\Dcand\to h^{\pm}h^{\mp}h^{\pm}$ decays to a negligible level.

Candidate \DandDsToPhiPi\ decays are reconstructed 
from three charged particles
originating from a single vertex.   
The particles are required 
to have $\chisqip>15$ and a scalar sum $\pt>2.8$\gevc.
The $\phi$ candidate is formed from a pair of 
oppositely charged particles that are identified as kaons
and have $\pt>0.25$\gevc.
The invariant mass of the $K^+K^-$ pair is required
to be within $20$\mevcc of the known $\phi$ mass~\cite{PDG2012}.
The bachelor pion is required to have $p>5$\gevc, $\pt>0.5$\gevc
and be identified as a pion.

Candidate \Dcand\ mesons in all decay modes
are required to have 
$\pt>1$\gevc, $\chisqip<9$
and vertex $\chi^2$ per degree of freedom less than $10$.
In addition, 
the \DToKsh\ (\ensuremath{\DandDsToPhiPi\xspace}) candidates
are required to have a 
vertex separation $\chi^2$ to the PV larger than $30$ ($125$),
a distance of closest approach of the decay products smaller
than $0.6$ ($0.5$)~mm,
and a cosine of the angle between the
\Dcand\ momentum and the vector between the PV and the \Dcand\
vertex greater than $0.999$.
The $\Dcand$ mass is required to be in the range 
$1.79 < m(\KS h^{\pm})< 2.03$\gevcc and
$1.805< m(K^+K^- \pi^{\pm})< 2.035$\gevcc for the \DToKsh\ 
and \DandDsToPhiPi\ decays, respectively.

\begin{figure}[tb]
  \begin{center}
    \subfigure{\includegraphics*[width=7.5cm]{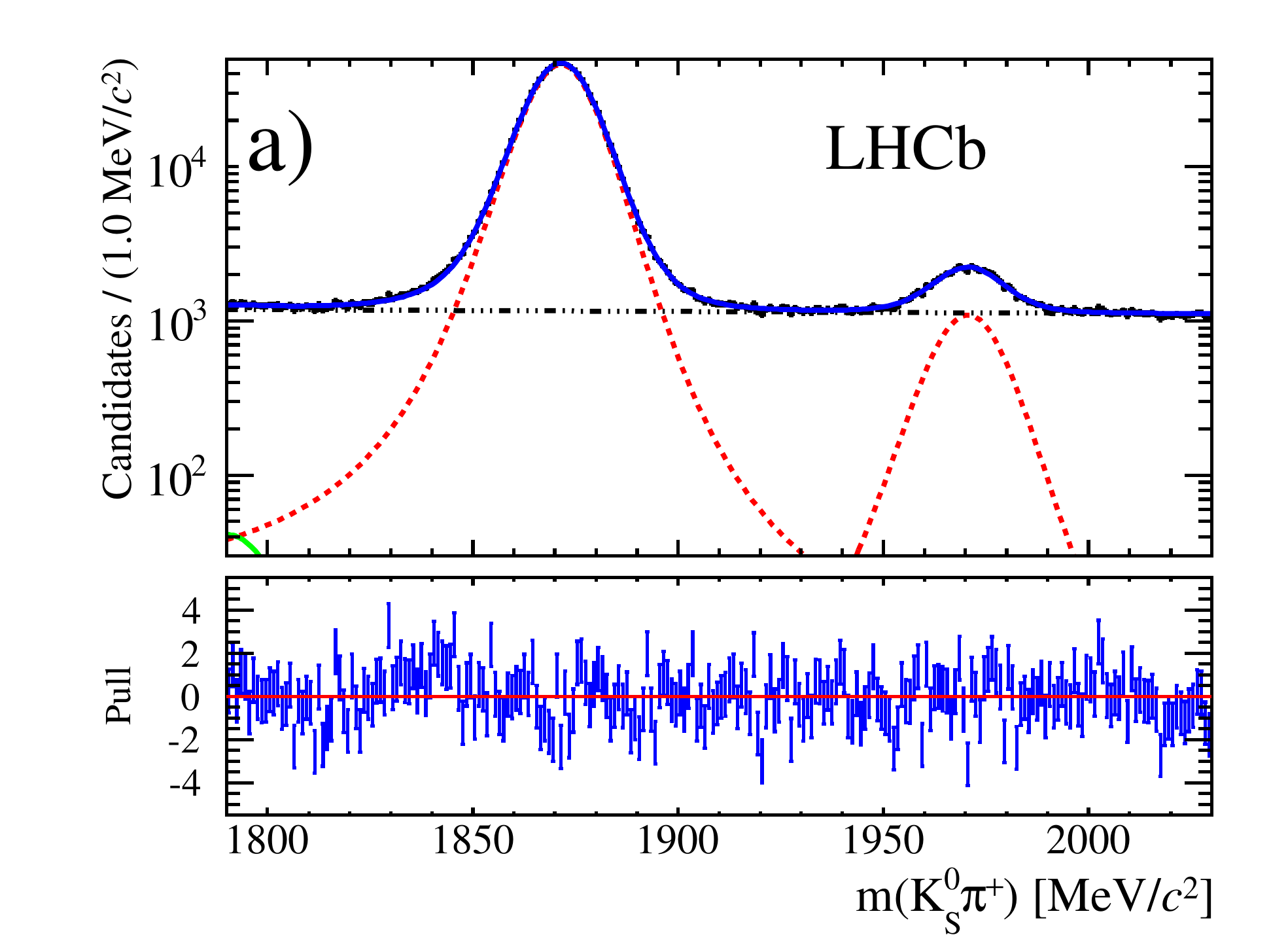}}
    \subfigure{\includegraphics*[width=7.5cm]{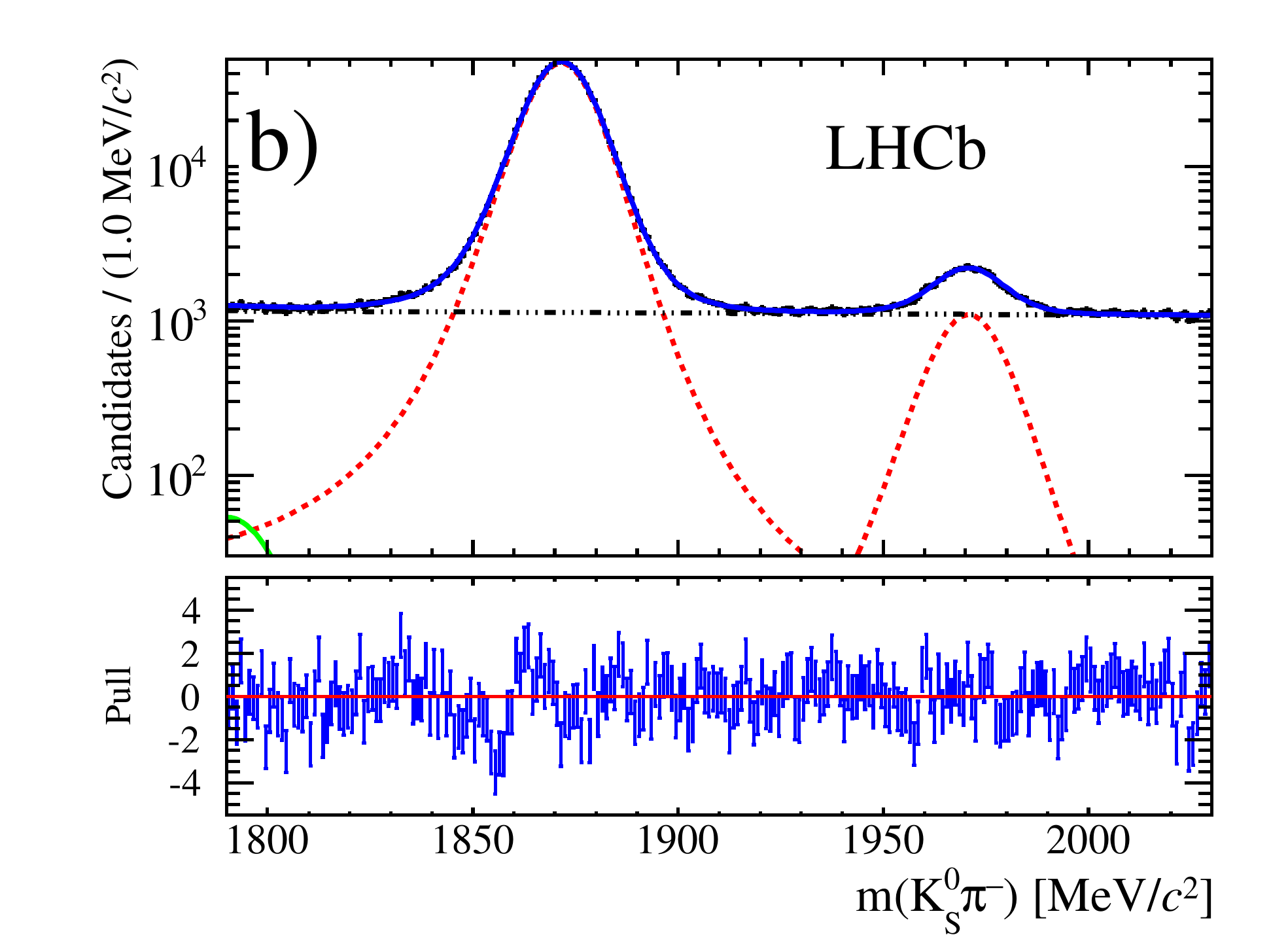}}
    \subfigure{\includegraphics*[width=7.5cm]{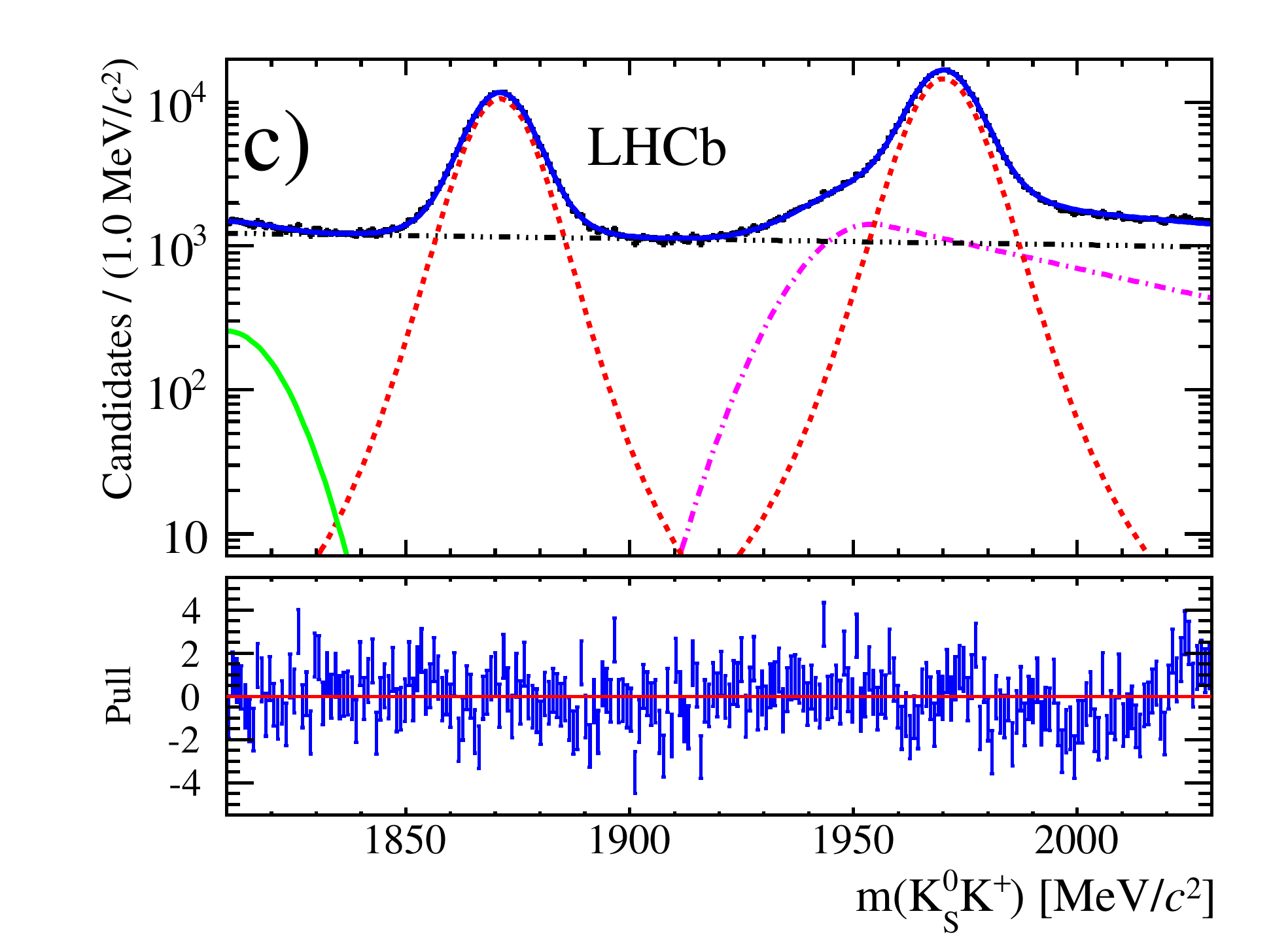}}
    \subfigure{\includegraphics*[width=7.5cm]{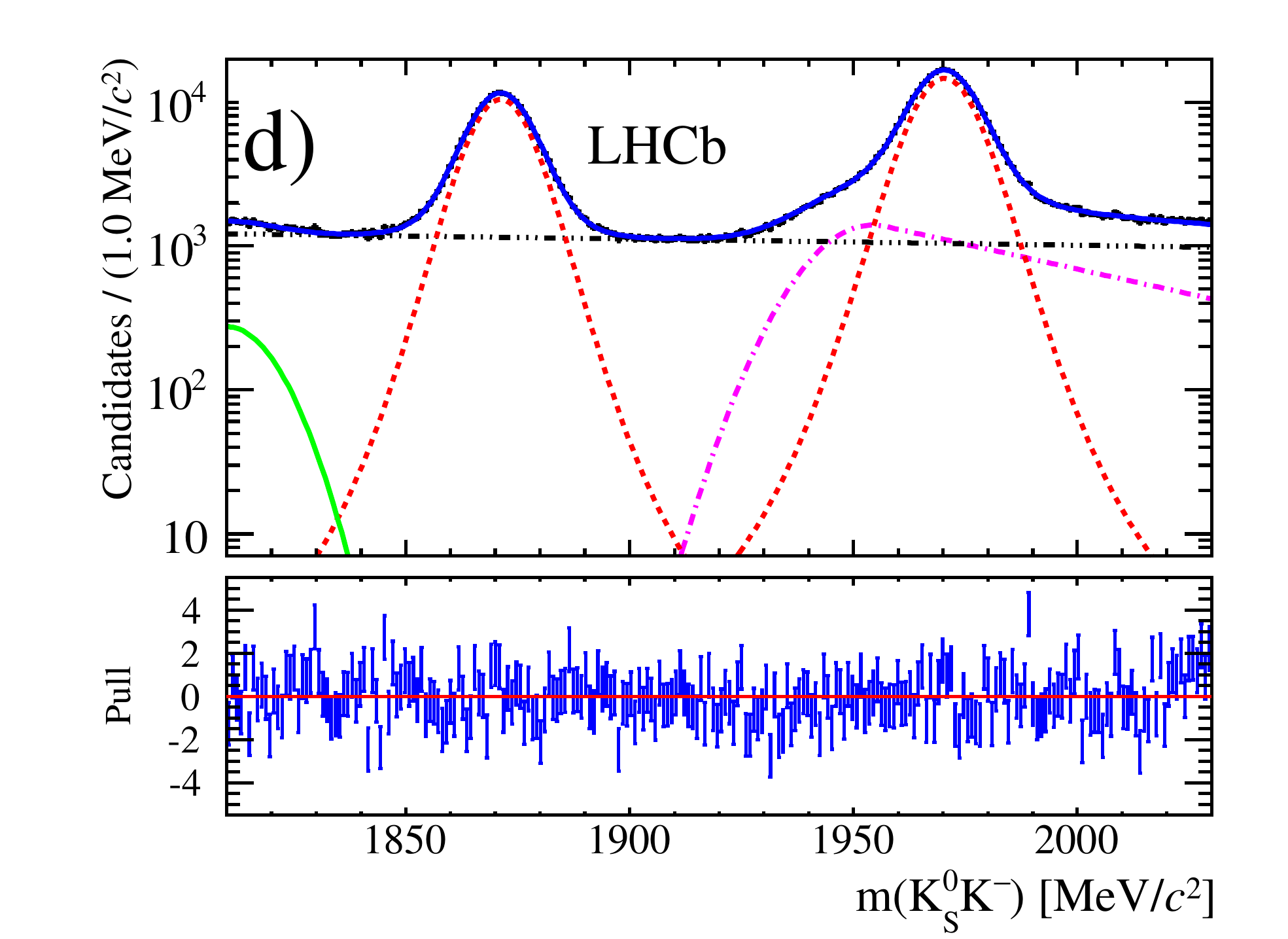}}
\caption{
\small{Invariant mass distributions for the 
a) \DcandToKsPip, 
b) \DcandToKsPim, 
c) \DcandToKsKp\ and 
d) \DcandToKsKm\ 
decay candidates for data taken in the 
magnetic polarity {\it Up} configuration
at $\sqrt{s}=8$~TeV.
The data are shown as black points and the total fit function 
by a blue line.
The contributions from the signal 
and the low-mass, cross-feed and combinatorial backgrounds
are indicated by red (dotted), green (full), magenta (dash-dotted) 
and black (multiple-dot-dashed) lines, respectively.
The bottom figures are the normalised residuals (pull) distributions.
}}
\label{fig:Baseline-MagUp-2012}
  \end{center}
\end{figure}

\begin{figure}[tb]
  \begin{center}
    \subfigure{\includegraphics*[width=7.5cm]{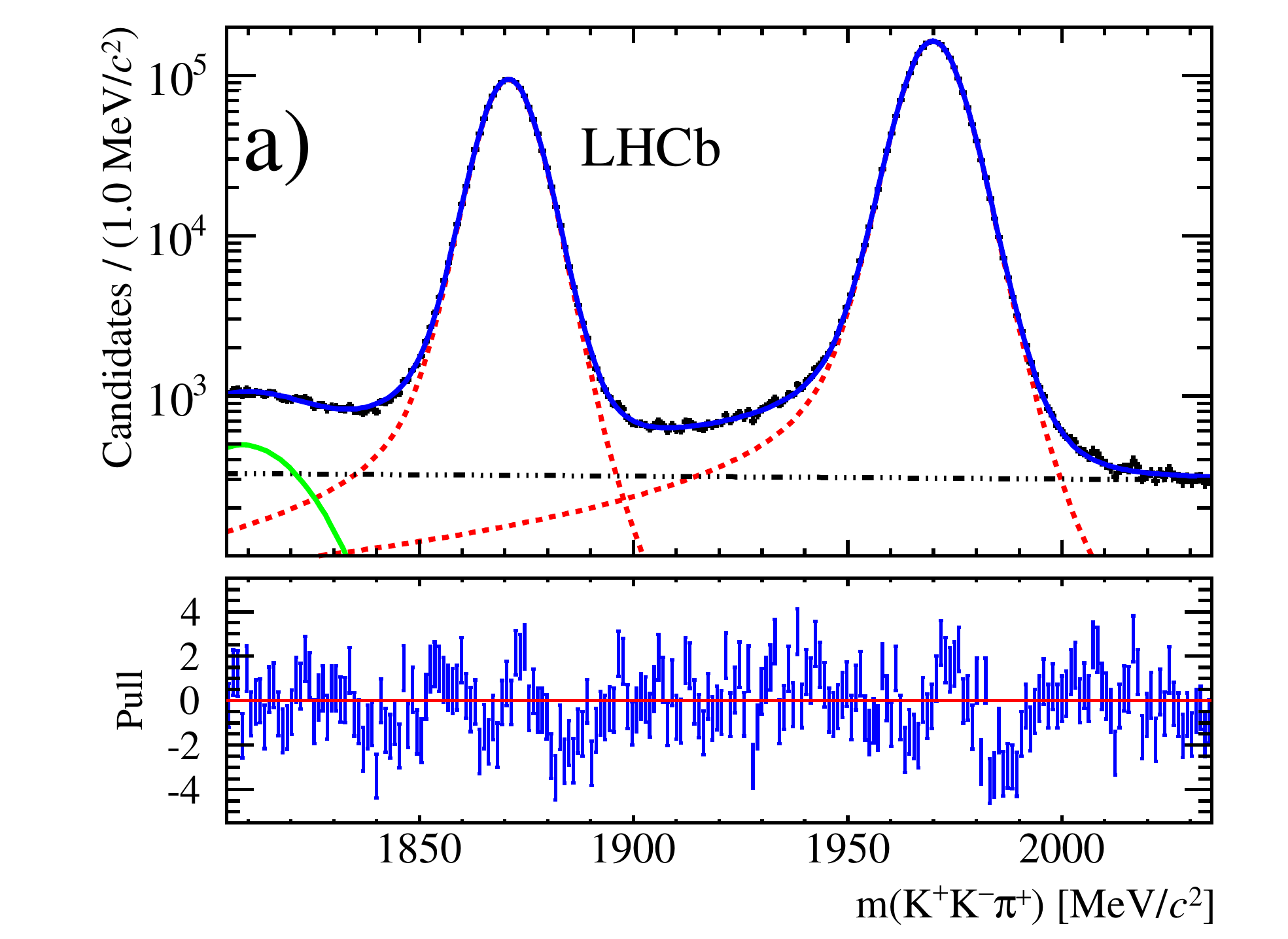}}
    \subfigure{\includegraphics*[width=7.5cm]{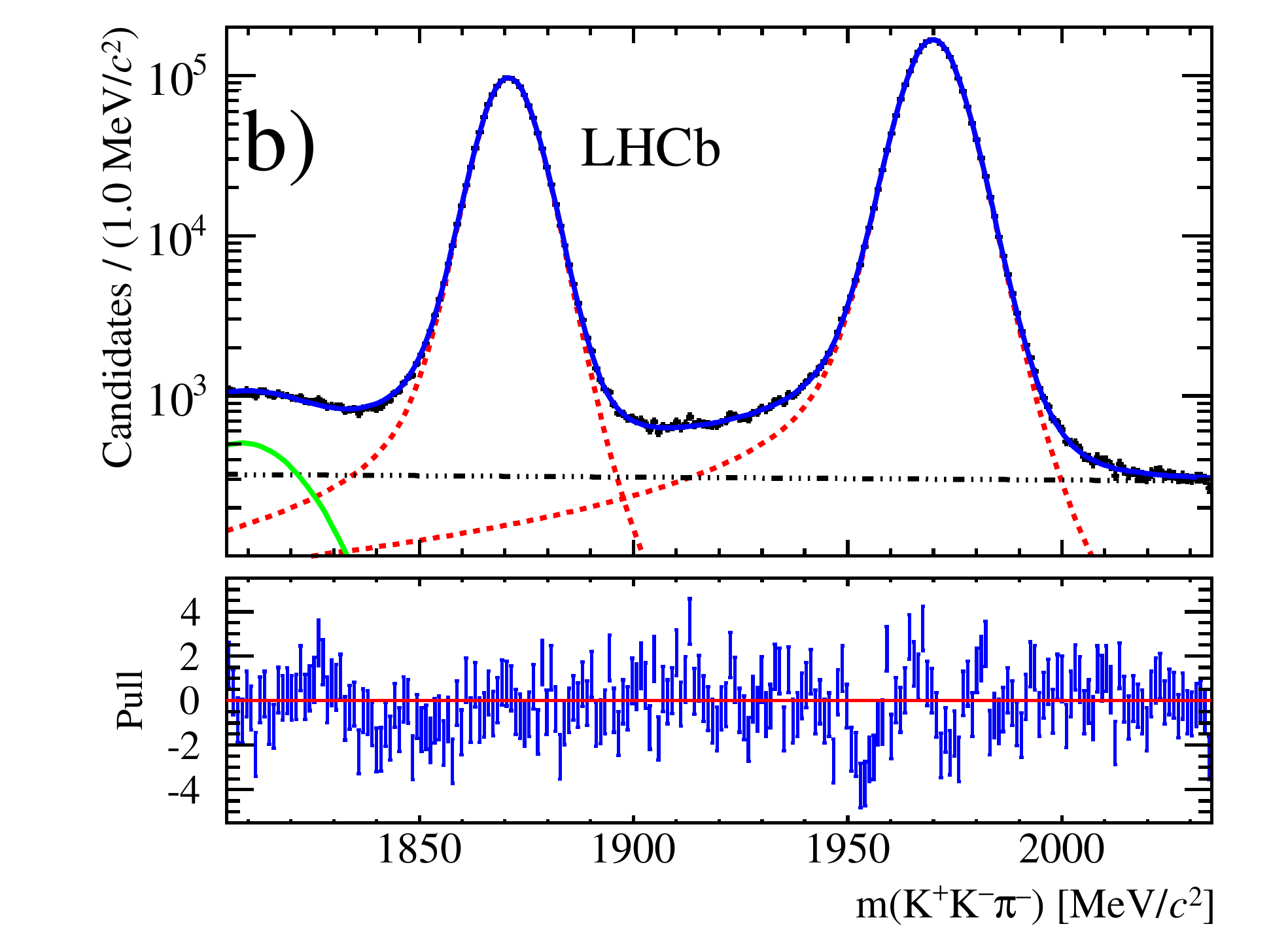}}
 \caption{
\small{Invariant mass distributions for the 
a) \DcandToPhiPip\ and
b) \DcandToPhiPim\ 
decay candidates for data taken in the 
magnet polarity {\it Up} configuration 
at $\sqrt{s}=8$~TeV.
The data are shown as black points and the total fit function
by a blue line.
The contributions from the signal and the low-mass and combinatorial backgrounds
are indicated by red (dotted), green (full) and black (multiple-dot-dashed) 
lines, respectively.
The bottom figures are the normalised residuals (pull) distributions.
}}
\label{fig:Baseline-MagUp-Control-Channels-2012}
  \end{center}
\end{figure}

Figures~\ref{fig:Baseline-MagUp-2012} 
and \ref{fig:Baseline-MagUp-Control-Channels-2012} 
show the mass distributions of
selected \DToKsh\ and \DandDsToPhiPi\ candidates for data taken 
in the magnet polarity {\it Up} configuration
at $\sqrt{s}=8$\tev.
The mass distributions for the magnet polarity {\it Down} configuration
are approximately equal.

Three categories of background contribute to the selected \Dcand\ candidates.
A {\it low-mass} background contributes at low \Dcand\ mass and
corresponds to decay modes 
such as $\Dpm\to\KS\pi^{\pm}\pi^0$ and
$\D_s^{\pm}\to K^{\mp}K^{\pm}\pi^{\pm}\pi^0$, 
where the $\pi^0$ is not reconstructed,
for \DToKsh\ and \DandDsToPhiPi\ decays, respectively.
A {\it cross-feed} background contributes to \DToKsh\ decays
and arises from \DToKshprime\ decays in which the bachelor pion (kaon) is
misidentified as a kaon (pion).
Simulation studies show that the misidentification 
of the bachelor pion in $D^{\pm}\to \KS\pi^{\pm}$ decays produces
a cross-feed background that extends under 
the \DsToKsK\  
signal peak, 
and that the bachelor kaon in \DsToKsK\
decays produces a small complementary cross-feed background
that extends under the $D^{\pm}\to \KS\pi^{\pm}$ signal peak.
A {\it combinatorial} background contribution is present in both
\DToKsh\ and \DandDsToPhiPi\ decay modes.
Background from $\Lz_c^{\pm}$ decays 
with a proton in the final state,
and \Dcand\ mesons originating from the decays of $b$~hadrons
are neglected in the fit and considered when assessing
systematic uncertainties. 

\section{Fit method}
\label{sec:Fit}

The yields and asymmetries 
for the \DandDsToKsPi, \DandDsToKsK,
and \DandDsToPhiPi\ signal channels
and the various backgrounds
are determined from a 
likelihood fit to the respective binned invariant mass distribution.
For each final state, 
the data are divided into four independent subsamples, 
according to magnet polarity and candidate charge,
and a simultaneous fit is performed.
The $\sqrt{s}=7$\tev and 8\tev data sets are fitted separately
to take into account background rate and data-taking conditions.

All signal and background mass shapes 
are determined using simulated data samples.
The \DToKsh\ signal shape is
described by the parametric function, 
\begin{equation}
f(m) \propto \exp\left[\frac{-(m-\mu)^2}{2\sigma^2+(m-\mu)^2\alpha_{L,R}}\right],
\label{eq:cruijff}
\end{equation}
which is parametrised by a mean $\mu$, width $\sigma$ 
and asymmetric low- and high-mass tail parameters,
$\alpha_L$ (for $m<\mu$) and $\alpha_R$ (for $m>\mu$), 
respectively. 
The means and widths of the four \Dcand\ signal peaks
are allowed to vary in the fit. 
All the \DandDsToKsPi\ signal peaks are described by 
two common $\alpha_L$ and $\alpha_R$ tail parameters,
whereas for the \DandDsToKsK\ signal peaks $\alpha_L$ and $\alpha_R$ 
are set to be equal and a single tail parameter is used. 
The widths and tail parameters are also common for the two magnet polarities.

The low-mass background is modelled by a Gaussian function 
with a fixed mean 
($1790$\mevcc and $1810$\mevcc for 
\DandDsToKsPi\ and \DandDsToKsK, respectively) 
and width ($10$\mevcc ), as determined from simulation.
The cross-feed components are described by a 
Crystal Ball function~\cite{Skwarnicki:1986xj} 
with tail parameters fixed to those obtained in the simulation.
Since the cross-feed contribution from \DsToKsK\ is very small compared
to the \DToKsPi\ signal,
the width and mean of this contribution are also taken
from simulation.
The cross-feed contribution from
\DToKsPi\ to \DsToKsK\ candidates
extends under the signal peak to low- and high-mass.
The mean and width of the Crystal Ball function are
allowed to vary in the fit
with a common width for the two magnet polarities.  
The combinatorial background is described by 
a linear term with a slope free to vary 
for all mass distributions. 

The \DandDsToPhiPi\ signal peaks are described by 
the sum of \Equation~(\ref{eq:cruijff})
and a Crystal Ball function.
The means and widths of the four \Dcand\ signal peaks
and a common Crystal Ball width
are allowed to vary in the fit.
In addition, five tail parameters are included in the fit.
These are $\alpha_L$ for the \Dpm\ and $D_s^{\pm}$ signal peaks
and a single offset 
$\Delta\alpha \equiv \alpha_L-\alpha_R$, 
and two Crystal Ball tail parameters. 
The widths and tail parameters are common for the two magnet polarities. 
The low-mass background is modelled with a Gaussian function and 
the combinatorial background is described by 
a linear term with a slope free to vary 
for all mass distributions.

To reduce any bias in the measured asymmetries due to
potential detection and production asymmetries arising from
the difference in the kinematic properties 
of the \Dcand\ or the bachelor hadron,
the \pt and $\eta$ distributions of the \Dcand\ candidate 
for the \DandDsToKsPi\ 
and \DandDsToPhiPi\ decay modes are weighted to be consistent 
with those of the \DandDsToKsK\ candidates.
To further reduce a potential bias due to a track detection asymmetry,
an unweighted average of the asymmetries measured using 
the two magnet polarity configurations is determined.

The total fitted signal yields for all decay modes and 
the measured and calculated \CP asymmetries
are summarised in 
\Table~\ref{tab:yields} and \Table~\ref{tab:asymmetries}, 
respectively.
Since the correlation between the measured asymmetries is negligible, 
the \CP asymmetries are calculated assuming they are uncorrelated.

\begin{table}[tb]
  \setlength{\tabcolsep}{0pt}
\caption{
  \label{tab:yields}
  \small{Signal yields.}}
  \begin{center}
    \begin{tabular}{lrl} \hline
Decay mode  & \multicolumn{2}{c}{Yield} \\ \hline
\DToKsPi\   & $4\,834\,440\,$ & $\pm \,2\,555$ \\
\DsToKsPi\  & $   120\,976\,$ & $\pm \,\phantom{2}\,692$ \\
\DToKsK\    & $1\,013\,516\,$ & $\pm \,1\,379$ \\
\DsToKsK\   & $1\,476\,980\,$ & $\pm \,2\,354$ \\
\DToPhiPi\  & $7\,020\,160\,$ & $\pm \,2\,739$ \\
\DsToPhiPi\ & $13\,144\,900\,$& $\pm \,3\,879$ \\ \hline     
    \end{tabular}
  \end{center}
\end{table}

\begin{table}[tb]
\caption{
  \label{tab:asymmetries}
  \small{Measured asymmetries (in \%) for the decay modes \DToKsPi, \DsToKsPi, 
    \DsToKsK\ and \DsToPhiPi\ and the calculated \CP asymmetries.
    The results are reported separately for
    $\sqrt{s}=7$~TeV and $\sqrt{s}=8$~TeV data and the 
    two magnetic polarities ({\it Up} and {\it Down}). 
    The combined results are given in the final column.
    The quoted uncertainties are statistical only.}}
  \begin{center}
    \begin{tabular}{l|cc|cc|c} \hline
                       & \multicolumn{2}{c}{$\sqrt{s}=7$~TeV} 
                       & \multicolumn{2}{c}{$\sqrt{s}=8$~TeV}  \\
Asymmetry              & {\it Up} & {\it Down} & {\it Up} & {\it Down} & Total \\ \hline
$\ACPmeas^{\DToKsPi}$  & $-1.04\pm 0.19$   
                       & $-0.74\pm 0.16$     
                       & $-0.88\pm 0.08$
                       & $-1.04\pm 0.08$     
                       & $-0.95\pm 0.05$        \\
$\ACPmeas^{\DsToKsPi}$ & $+2.55\pm 1.34$   
                       & $-0.56\pm 1.09$     
                       & $-0.46\pm 0.78$   
                       & $-0.66\pm 0.77$     
                       & $-0.15\pm 0.46$        \\
$\ACPmeas^{\DToKsK}$   & $-0.47\pm 0.59$   
                       & $-0.23\pm 0.50$     
                       & $-0.11\pm 0.32$   
                       & $+0.38\pm 0.31$      
                       & $+0.01\pm 0.19$        \\
$\ACPmeas^{\DsToKsK}$  & $+0.28\pm 0.34$   
                       & $+0.84\pm 0.28$     
                       & $-0.69\pm 0.18$   
                       & $+1.02\pm 0.17$     
                       & $+0.27\pm 0.11$        \\
$\ACPmeas^{\DsToPhiPi}$& $-1.02\pm 0.09$   
                       & $+0.24\pm 0.07$     
                       & $-0.71\pm 0.05$   
                       & $-0.48\pm 0.05$     
                       & $-0.41\pm 0.05$        \\ \hline
$\ACPDD        $       & $+2.71\pm 1.46$   
                       & $-1.04\pm 1.18$     
                       & $+0.86\pm 0.82$   
                       & $-0.39\pm 0.81$     
                       & $+0.41\pm 0.49$         \\ 
$\ACP^{\DToKsK}$       & $-0.80\pm 0.53$   
                       & $-0.17\pm 0.44$     
                       & $+0.69\pm 0.27$   
                       & $-0.14\pm 0.27$     
                       & $+0.03\pm 0.17$        \\ 
$\ACP^{\DsToKsPi}$     & $+3.51\pm 1.35$   
                       & $-0.87\pm 1.09$     
                       & $+0.17\pm 0.78$   
                       & $-0.25\pm 0.77$     
                       & $+0.38\pm 0.46$        \\ \hline
    \end{tabular}
  \end{center}
\end{table}

\section{Systematic uncertainties}
\label{sec:Systematics}

The values of the \CP asymmetries $\ACPDD$,
$\ACP^{\DToKsK}$ and $\ACP^{\DsToKsPi}$
are subject to several sources of systematic uncertainty
arising from the fitting procedure, treatment of the backgrounds,
and trigger- and detector-related effects.
A summary of the contributions 
to the systematic uncertainties is given in \Table~\ref{tab:systematics}.
\begin{table}[tb]
\caption{
  \label{tab:systematics}
  \small{Systematic uncertainties (absolute values in \%) 
    on the \CP asymmetries for
    $\sqrt{s}=7$ and $8$~TeV data. The total systematic uncertainty
    is the sum in quadrature of the individual contributions.}}
  \begin{center}
    \begin{tabular}{l|ccc|ccc} \hline
                  & \multicolumn{3}{c}{$\sqrt{s}=7$~TeV}
                  & \multicolumn{3}{c}{$\sqrt{s}=8$~TeV}             \\
Source            & $\ACPDD$ & $\ACP^{\DToKsK}$ & $\ACP^{\DsToKsPi}$  
                  & $\ACPDD$ & $\ACP^{\DToKsK}$ & $\ACP^{\DsToKsPi}$ \\ \hline
Fit procedure     & $0.14$   & $0.09$           & $0.11$                   
                  &  $0.07$  & $0.05$           & $0.01$             \\
Cross-feed bkgd.  & $0.03$   & $0.01$           & $0.02$                   
                  &  $0.01$  & $-$              & $0.01$             \\
Non-prompt charm  & $0.01$   & $-$              & $-$                   
                  &  $0.01$  & $-$              & $-$                \\
Kinematic weighting        & $0.08$   & $0.06$           & $0.13$                   
                  &  $0.05$  & $0.07$           & $0.12$             \\
Kinematic region   & $0.10$   & $0.06$           & $0.04$                   
                  &  $0.19$  & $0.02$           & $0.17$             \\
Trigger           & $0.13$   & $0.13$           & $0.07$                    
                  & $0.17$   & $0.17$           & $0.09$             \\
$\Kz$ asymmetry & $0.03$      & $0.02$           & $0.02$                     
                    &  $0.04$     & $0.02$           & $0.02$             \\ \hline

Total             & $0.23$   & $0.18$           & $0.19$                     
                  &  $0.27$  & $0.19$           & $0.22$             \\ \hline
    \end{tabular}
  \end{center}
\end{table}

The systematic uncertainty due to the fit procedure 
is evaluated by replacing the description of the 
$\DToKsh$ and $\DandDsToPhiPi$ signal, 
combinatorial background
and low-mass background
in the fit with alternative parameterizations. 
The systematic uncertainty is calculated by comparing the 
asymmetries after each change in the fit function
to those obtained without the modification. 
The overall systematic uncertainty due to the fit procedure
is calculated assuming that the
individual contributions are entirely correlated.

The systematic uncertainty due to the $\DsToKsK$ cross-feed in the 
$\DcandToKsPi$ fit is determined by repeating the fit with
the cross-feed component yields fixed to those 
from an estimation based on
particle identification efficiencies determined from
a large sample of $D^{*\pm}\to D \pi^{\pm}$ decays,
where $D$ is a $\Dz$ or $\Dzb$ meson~\cite{LHCb-PROC-2011-008}.
In the $\DcandToKsK$ fit, 
the $\DToKsPi$ cross-feed shape tail parameters are allowed to vary.
The systematic uncertainty is taken as the shift in the central values
of the \CP asymmetries.

The systematic uncertainty due to the presence of
charm backgrounds, 
such as 
$\Lz_c^{\pm}\to \Lz^0 h^{\pm}$ and
$\Lz_c^{\pm}\to \KS p$,
which have a proton in the final state,
is investigated by applying a proton identification veto
on all final state tracks in the $\DToKsh$ data sample.
The effect is to reduce the total number of $\DToKsh$ 
candidates, 
without a significant shift in the asymmetries.
This source of systematic uncertainty is therefore considered
negligible. 

In the selection of \Dcand\ candidates,
the \chisqip requirement on the \Dcand\ 
removes the majority of background from secondary 
\Dcand\ mesons originating from 
the decay of a $b$ hadron.
The remaining secondary \Dcand\ mesons 
may introduce a bias in the measured \CP asymmetries due 
to a difference in the 
production asymmetries for $b$ hadrons and \Dcand\ mesons. 
In order to investigate this bias, 
the \Dcand\ production asymmetries in \Equation~(\ref{e:ACPDToKsh}) 
for \DToKsh\ decays, and similarly for \DandDsToPhiPi\ decays,
are modified using
\begin{align}
\label{e:ACPprodcorr}
\ACPprod^{\Dcand} (corr) = \dfrac{\ACPprod^{\Dcand}+f \ACPprod^{B}}{1+f},
\end{align}
where $f$ is the fraction of secondary \Dcand\ candidates in
a particular decay channel
and $\ACPprod^{B}$ is the corresponding $b$-hadron production asymmetry.
The fraction $f$ is estimated from the
measured 
$\Dpm$, 
$\D_s^{\pm}$ 
and $b$ hadron 
inclusive cross-sections~\cite{LHCb-PAPER-2012-041,LHCb-PAPER-2010-002},
the inclusive branching fractions
$\BR(b\to D^{\pm}X)$ and $\BR(b\to D^{\pm}_s X)$,
where $X$ corresponds to any other 
particles in the final state~\cite{PDG2012},
the exclusive branching fractions 
$\BR(\DToKsh )$ and $\BR(\DandDsToPhiPi)$~\cite{PDG2012},
and the efficiencies estimated from simulation.
The resulting values of $f$ lie in the range $1.3-3.2\%$.
The $b$-hadron production asymmetry $\ACPprod^{B}$ 
is taken to be $(-1.5\pm 1.3)\%$,
consistent with measurements of the $B^+$ and $B^0$ 
production asymmetries in $pp$ collisions 
in the forward region~\cite{LHCb-PAPER-2012-001}.
The effect of the uncertainty on $\ACPprod^{B}$ is negligible.
The systematic uncertainty is evaluated by using the modified
\Dcand\ production asymmetries from \Equation~(\ref{e:ACPprodcorr})
for each of the decay modes and recalculating the \CP asymmetries.

The effect on the \CP asymmetries of weighting the 
\DcandToKsPi\ and \DandDsToPhiPi\ candidates
using the \Dcand\ kinematic distributions
compared to the unweighted results is assigned as a systematic uncertainty.
The effect of the weighting procedure on 
the bachelor hadron kinematic distributions 
is also investigated by comparing the 
bachelor \pt and $\eta$ distributions before and after weighting.
The results show excellent agreement and no further systematic uncertainty
is assigned.  

Due to a small intrinsic left-right detection asymmetry,
for a given magnet polarity,
an excess of 
either positively or negatively charged 
bachelor hadrons is detected at large $\eta$ and small $p$, 
where $p$ is the component of momentum 
parallel to the LHCb beam-axis~\cite{LHCb-PAPER-2012-009}.
This excess leads to charge asymmetries, 
which may not completely cancel in the analysis
when the average of the 
{\it Up} and {\it Down} magnet polarity asymmetries is calculated.
To investigate this effect,
\Dcand\ candidates,
whose bachelor hadron falls within the 
above kinematic region,
are removed and the resulting asymmetries compared to those without the
selection criterion applied.
The kinematic region excluded is the same as that used
in \References~\cite{LHCb-PAPER-2012-009, LHCb-PAPER-2012-026} 
and removes $\sim 3\%$ of the \Dcand\ candidates.
The difference between the asymmetries is taken to be
the systematic uncertainty. 

Detector related systematic uncertainties may also arise from 
the variation of operating conditions between data-taking
periods,
and data not taken concurrently with the two magnet polarities.
A consistency check is therefore performed
by dividing the data
into 12 subsamples with similar size, 
corresponding to data-taking periods and magnet polarity changes,
and the analysis is repeated for each subsample.
The asymmetries obtained are consistent and no further 
systematic uncertainty is assigned. 

Potential trigger biases are studied using a large sample of
\DToKPiPi\ decays with the \DandDsToPhiPi\ selection criteria applied.
The data are divided into subsamples,
corresponding to various 
hardware trigger configurations,
and the asymmetries for the individual subsamples measured.
A systematic uncertainty is assigned, 
which corresponds to the maximum
deviation of a \CP asymmetry from a single subsample
compared to the mean asymmetry from all subsamples,
assuming there is no cancellation when
the \CP asymmetries 
are remeasured. 

In \DToKsh\ decays,
the \KS meson originates from the production
of a neutral kaon flavour eigenstate
($\Kz$ or $\Kzb$) in the decay of the $\Dcand$ meson.
The neutral kaon state evolves, 
via mixing and \CP violation,
and interacts with the detector material 
creating an asymmetry 
in the reconstruction before decaying. 
The overall effect is estimated using simulation,
as described in \Reference~\cite{LHCb-PAPER-2014-013}, 
and a correction is applied to the calculated asymmetries
as shown in \Equations~(\ref{e:ACPDDest})-(\ref{e:ACPDsToKsPi}).
The full uncertainty of the estimated effect 
is assigned as a systematic uncertainty.

\section{Results and summary}
\label{sec:Results}

A search for \CP violation in \DToKsK\ and \DsToKsPi\ decays is performed
using a data sample of $pp$ collisions,
corresponding to an integrated luminosity 
of $3$\invfb at centre-of-mass energies of 7~TeV ($1$\invfb)
and 8~TeV ($2$\invfb),
recorded by the LHCb experiment.
The results for the two centre-of-mass energies are combined
using the method described in \Reference~\cite{Lyons:1988rp},
assuming all the systematic uncertainties are correlated.
The individual \CP-violating asymmetries are measured to be  
\begin{align}
\ACP^{\DToKsK} = (+0.03\pm 0.17\pm 0.14)\% \nonumber
\end{align}
and
\begin{align}
\ACP^{\DsToKsPi} = (+0.38\pm 0.46\pm 0.17)\%, \nonumber
\end{align}
assuming that \CP violation in the Cabibbo-favoured decay is
negligible.
The measurements are consistent with previous 
results~\cite{Ko:2012uh,LHCb-PAPER-2012-052}, 
and $\ACP^{\DsToKsPi}$ supersedes the result 
reported in \Reference~\cite{LHCb-PAPER-2012-052},
which used a subsample of the present data.

A combination of the measured asymmetries for the four decay modes
\DcandToKsK\ and \DcandToKsPi\ gives the sum
\begin{align}
\ACP^{\DToKsK}+\ACP^{\DsToKsPi} = (+0.41\pm 0.49\pm 0.26)\%, \nonumber
\end{align}
and provides a measurement that is largely insensitive to production
and instrumental asymmetries.  
In all cases,
the first uncertainties are statistical and the second are systematic.
The results represent the most precise measurements of these quantities
to date and show no evidence for \CP violation.

\section*{Acknowledgements}
 
\noindent We express our gratitude to our colleagues in the CERN
accelerator departments for the excellent performance of the LHC. We
thank the technical and administrative staff at the LHCb
institutes. We acknowledge support from CERN and from the national
agencies: CAPES, CNPq, FAPERJ and FINEP (Brazil); NSFC (China);
CNRS/IN2P3 (France); BMBF, DFG, HGF and MPG (Germany); SFI (Ireland); INFN (Italy); 
FOM and NWO (The Netherlands); MNiSW and NCN (Poland); MEN/IFA (Romania); 
MinES and FANO (Russia); MinECo (Spain); SNSF and SER (Switzerland); 
NASU (Ukraine); STFC (United Kingdom); NSF (USA).
The Tier1 computing centres are supported by IN2P3 (France), KIT and BMBF 
(Germany), INFN (Italy), NWO and SURF (The Netherlands), PIC (Spain), GridPP 
(United Kingdom).
We are indebted to the communities behind the multiple open 
source software packages on which we depend. We are also thankful for the 
computing resources and the access to software R\&D tools provided by Yandex LLC (Russia).
Individual groups or members have received support from 
EPLANET, Marie Sk\l{}odowska-Curie Actions and ERC (European Union), 
Conseil g\'{e}n\'{e}ral de Haute-Savoie, Labex ENIGMASS and OCEVU, 
R\'{e}gion Auvergne (France), RFBR (Russia), XuntaGal and GENCAT (Spain), Royal Society and Royal
Commission for the Exhibition of 1851 (United Kingdom).



\addcontentsline{toc}{section}{References}
\setboolean{inbibliography}{true}
\bibliographystyle{LHCb}
\bibliography{DToKsh,main,LHCb-PAPER,LHCb-CONF,LHCb-DP}

\end{document}